\documentclass[preprint]{jpsj3}
\usepackage{graphicx,bm}

\title{Effect of the RuO$_6$ octahedron rotation at the Sr$_2$RuO$_4$ surface on topological property}

\author{Yoshiki Imai$^1$\thanks{imai@phy.saitama-u.ac.jp}, Katsunori Wakabayashi$^2$, and Manfred Sigrist$^3$}
\inst{$^1$Department of Physics, Saitama University, Saitama, 338-8570, Japan \\
$^2$International Center for Materials Nanoarchitectonics (WPI-MANA), National Institute for Materials Science (NIMS),
Tsukuba 305-0044, Japan \\
$^3$Theoretische Physik, ETH-H\"onggerberg, CH-8093 Z\"urich, Switzerland} 

\abst{
We investigate the rotation effect of the RuO$_6$ octahedron around the $c$ axis on the topological and transport properties near the surface of the spin-triplet superconductor Sr$_2$RuO$_4$. While the Fermi level of bulk Sr$_2$RuO$_4$ is near the Lifshitz transition, the RuO$_6$ rotation realized near the surface leads to the change of the Fermi surface topology.  The edge current resulting from the time-reversal symmetry breaking in the chiral $p$-wave phase with fully opened excitation gap is less affected around Lifshitz transition. The topological property and the edge state are sensitive to the rotation angle and the amplitude of the nearest neighbor interaction, and the superconducting gap is strongly reduced in the larger next nearest neighbor interaction region. 
Although the edge state in Sr$_2$RuO$_4$ is topologically protected, it is not robust to the disorder such as impurity or defect. 
}

\begin{document}
\maketitle
\section{Introduction}
The transition metal compound Sr$_2$RuO$_4$ has attracted much interest due to the discovery of the unconventional superconductivity~\cite{maen94,mack03,maen12}. 
$\mu$SR experiments suggest a time-reversal symmetry breaking superconducting state~\cite{luke98} and the Knight-shift in the NMR experiments is compatible with spin-triplet Cooper pairing~\cite{ishi98}.  
The leading candidate of the superconducting order parameter has the so-called chiral $p$-wave symmetry, represented as
\begin{eqnarray}
\bm d=\Delta_0\hat{z}(k_x\pm ik_y), 
\end{eqnarray}
which is the two-dimensional analog of the Anderson-Brinkman-Morel (ABM) state of $^3$He superfluid~\cite{rice95}. 
There exists a full energy gap and orbital angular momentum $L_z=\pm 1$ of the Cooper pairs along the $z$ axis. 
The two angular momentum states, $k_x \pm ik_y$, are degenerate, which can lead to the formation of domains. 
Both domain walls and surfaces of the material host the local subgap states. Although the spontaneous appearance of supercurrents  
is expected based on theoretical arguments~\cite{volo85,mats99,laub00}, scanning Hall probe and scanning SQUID microscopy experiments give negative results~\cite{liu03,kash11,tame03}. 
 
Sr$_2$RuO$_4$ has a K$_2$NiF$_4$-type lattice structure whose space group belongs to $I4/mmm$ and shows strong two-dimensional anisotropy. It is well known that the low-energy electronic properties are dominated by the Ru 4d $t_{2g}$-orbitals, which generate three cylindrical  Fermi surfaces, the $\alpha$, $\beta$ and $\gamma$ Fermi surfaces~\cite{mazi97}. While the $\alpha$ and $\beta$ bands consist of the Ru $d_{yz}$ and $d_{zx}$ orbitals mainly and have essentially one-dimensional hole-like and electron-like characters, respectively, the $\gamma$ band originates from the $d_{xy}$ orbital and has two-dimensional electron-like structure. 

In previous papers, we discussed the magnetic and transport properties near the edges and the topological nature by means of the multi-band tight-binding model to the ribbon-shaped system~\cite{imai12,imai13}. 
 In the chiral $p$-wave phase, in addition to the supercurrent near the edges due to the time-reversal symmetry breaking, the spin-orbit interaction generates also a spin current resulting from the $\alpha$-$\beta$ bands. 
Since there exists the almost flat subgap state originating from the $\alpha$-$\beta$ bands at low energy region, the weak repulsive interaction easily gives rise to the spin polarization near the edges. The magnetic fields generated from the charge current and the spin polarization have similar amplitudes with opposite sign, which leads to the suppression of the spontaneous magnetic field. This is a possible way to explain that the discrepancy between the theoretical and the experimental studies concerning the presence of the edge current. 

On the other hand, the topological property mainly depends on the two-dimensional $\gamma$ band. While the Chern number of the two ($\alpha$-$\beta$) band model vanishes because the $\alpha$-$\beta$ bands have electron and hole characters, respectively, which yields a cancellation of the net Chern number, it has non-zero value in the three-band model which indicates that the $\gamma$ band is responsible for the topological property in Sr$_2$RuO$_4$.  
However, the $\gamma$ band is near the Lifshitz transition, which may be sensitive to the surface state. 
Early Angle-resolved Photoemission Spectroscopy (ARPES) results indicated a hole-like $\gamma$ band~\cite{yoko96}, while later experiments confirmed an electron-like Fermi surface, consistent with de Haas-van Alphen measurements~\cite{dama00}. 

Recently, Wang {\it et al.}~\cite{wang13} studied the pairing mechanism of Sr$_2$RuO$_4$ by means of the functional renormalization group technique, so that the superconducting gap in the chiral $p$-wave phase becomes smaller due to the contribution of the next nearest neighbor interactions, which leads to the weakness of the topological superconducting phase against disorder. 

The surface reconstruction that doubles the unit cell in Sr$_2$RuO$_4$, affecting particularly the $d_{xy}$ orbitals, has been reported, which complicates
the topology of the superconducting phase on the $\gamma$ band. Near the surface of Sr$_2$RuO$_4$ the rotation of the RuO$_6$ octahedron along the $c$ axis occurs with the doubling of the unit cell with $p4gm$ plane group symmetry, and the rotation angle $\theta$ is $9\pm 3^\circ$~\cite{matz00} and $7.46^\circ$~\cite{veen13} at the surface. 
The rotation effect reduces the hopping  amplitude between $d_{xy}$ orbitals which may affect the physical properties. 

In this paper, we investigate the effect of the surface reconstruction on the topological and transport properties by using the density functional theory and the lattice fermion model with the $\gamma$ band and the attractive interaction between the nearest and the next nearest neighbor interactions. We discuss the interplay between the doubling effect of unit cell and next nearest neighbor interaction. 

This paper is organized as follows. The detailed electronic structure is obtained by the density functional theory in next section. We construct the effective Hamiltonian with lattice fermion model, and discuss the transport and topological properties in Sec. 3, Summary and discussions are given in Sec. 4.

\section{{\it Ab initio} electronic structure of Sr$_2$RuO$_4$ near the surface}
In this section, we investigate the realistic electronic structure of tetragonal and the doubling of unite cell systems of Sr$_2$RuO$_4$ by means of the WIEN2k package~\cite{blah01} with density functional theory (DFT) based on the full-potential linearized augmented plane-wave (LAPW) method where the calculations are performed using the generalized gradient approximation (GGA)~\cite{perd96}. 
For the lattice constants the same values of the tetragonal structure (space group $I4/mmm$)~\cite{huan94} are employed.  
The muffin-tin (MT) sphere radius $R_{\rm MT}$ is give by $1.95$ for Ru, $2.3$ for Sr and $1.68$ for O, respectively. 
$R_{\rm MT}K_{\rm MAX}=7$ where $K_{\rm MAX}$ is amplitude of the largest $K$ vector in plane wave expansion. 

The similar ruthenium compound Sr$_3$Ru$_2$O$_7$ which shows no ordering down to low temperature region and the Fermi liquid behavior~\cite{huan98} has the rotation of the RuO$_6$ octahedron around the $c$ axis in the whole system~\cite{shak00}. 
Although in contrast to Sr$_3$Ru$_2$O$_7$ case, the rotation of the RuO$_6$ octahedron in Sr$_2$RuO$_4$ appears only near the surface,  we employ the lattice structure with the rotation of the RuO$_6$ octahedron not only at the surface but also bulk material, for simplicity. 
Since Sr$_2$RuO$_4$ has the strong two-dimensional anisotropy, the electronic state has also two-dimensional character and is almost independent of that of other layers. Thus the surface electronic state can be captured within this treatment.  

\begin{figure}[tb]
\includegraphics[width=90mm]{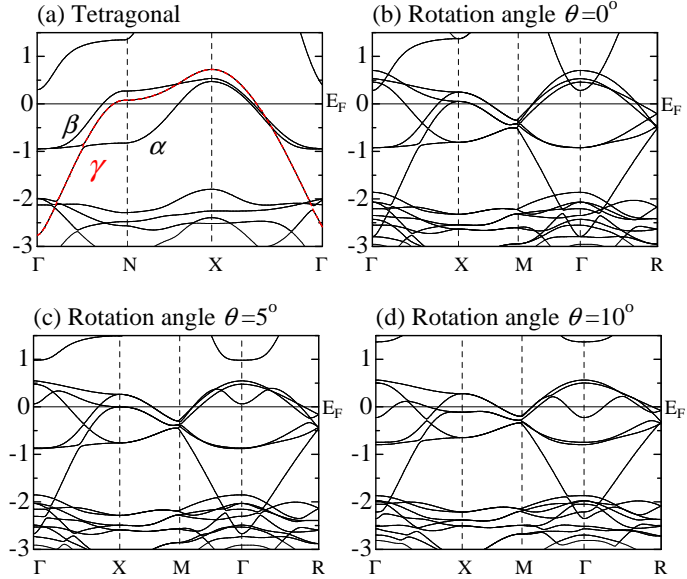}
\caption{(Color online) Band structures obtained from first principles calculation in the normal phase; (a) Tetragonal lattice (space group $I4/mmm$). The dashed line stands for the $\gamma$ band. (b)-(d)  Doubling of the unit cell with the rotation angle $\theta=0$, $\theta=5^\circ$ and $\theta=10^\circ$ around $c$ axis.  }
\label{band_LDA}
\end{figure}
Figure \ref{band_LDA} shows the band structures in the normal state from the first principles calculation for several choices of the rotation angle around the $c$ axis. 
The result of the rotation angle $\theta=0$ (Fig. \ref{band_LDA} (b)) is identical to that of the tetragonal lattice with space group $I4/mmm$ (Fig. \ref{band_LDA} (a)), and N point in Fig. \ref{band_LDA} (a) corresponds to X 
point in Fig. \ref{band_LDA} (b).
Compared with the number of the bands in Fig. \ref{band_LDA} (a), there are twice the number of the bands in Fig. \ref{band_LDA} (b)-(d) due to the doubling of the unit cell.  
Our result is consistent with the result from Veenstra et al. \cite{veen13}. 
 
In Fig. \ref{band_LDA} (a) near N point, the $\gamma$ band is slightly above the Fermi level where the $\gamma$ band lies near the van Hove point (N point).  
 With increasing the rotation angle $\theta$, the topmost position of the $\gamma$ band near N point shifts to lower energy region, and touches the Fermi level at $\theta\sim 5^\circ$ and is below the Fermi level at $\theta=10^\circ$. 
In Figs. \ref{band_LDA} (b)-(d) these band structures along $\Gamma$ to R are almost similar to those along $\Gamma$ to M, which indicates that the strong two-dimensionality remains even with the finite rotation of the RuO$_6$ octahedron. 
Note that the obtained band for $\theta=10^\circ$ has an electron-like pocket near $\Gamma$ point, which contains the component of Ru $4d$ $e_g$ orbitals and is especially similar to that of Sr$_3$Ru$_2$O$_7$~\cite{sing01}. 
 This result indicates that the Lifshitz transition occurs near the surface due to the rotation of the RuO$_6$ octahedron where the topological property may be different from the bulk one because the topological number may be considered separately for each layer due to the strong two-dimensionality.  
 
 While the rotation of the RuO$_6$ octahedron does not affect the Fermi surface topology of the $\alpha$-$\beta$ band (and its folded one), that of the $\gamma$ band is sensitive to the rotation angle.  
In the next section, by means of the lattice fermion model which describes the $\gamma$ band, we investigate the topological and transport properties systematically in the spin-triplet superconducting phase. 
 
\section{Study with Lattice Fermion Model}

\subsection{Effective model}
Lattice fermion models are suitable for the description of physical properties for many strongly correlated materials.  
Thus we construct the effective model with the lattice fermion model to discuss the effect of rotation of the RuO$_6$ octahedron on the physical property in the superconducting phase. 
The model consists of the kinetic energy and the interaction terms. 

Topological aspects are closely connected with the presence of surface states following from the bulk-edge correspondence~\cite{wen92,hats93}.  
\begin{figure}[t]
\includegraphics[width=70mm]{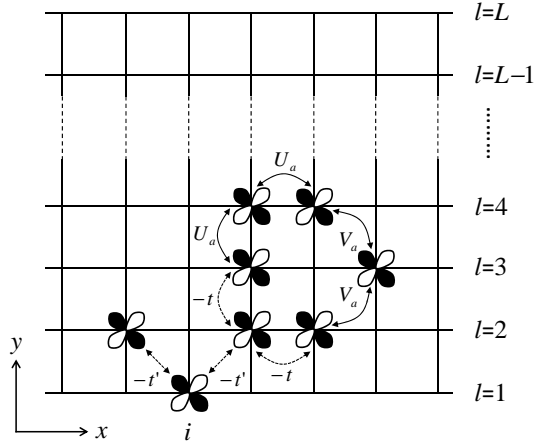}
\caption{Lattice structure with the number of legs $L$.  $t$ ($t'$) stands for the hopping amplitude between nearest (next nearest) neighbor lattice sites. $U_a$ ($V_a$) represents the attractive interaction between nearest (next nearest) neighbor lattice sites. }
\label{lattice}
\end{figure}
Thus we employ the ladder-type system which enables us to access the properties near the edges more easily,  
where open boundary conditions in $y$-direction, generating edge states, are taken into account. 
Figure \ref{lattice} shows the lattice ribbon structure with $L$ legs, leading to edges at $l=1$ and $l=L$.  
Assuming periodic boundary conditions between the leg $l=1$ and $l=L$ would yield a two-dimensional bulk system. 

The kinetic energy is composed of the two-dimensional tight-binding model describing the $\gamma$ band resulting from the $d_{xy}$ orbitals, since the topological property is independent of the presence of  the $\alpha$-$\beta$ band~\cite{imai13}.
The $d_{xy}$ orbitals, whose structure suggests the hopping matrix elements $t$ ($t'$) and the attractive interaction $U_a$ ($V_a$) between nearest neighbor (next nearest neighbor) sites. The latter is introduced to generate the superconductivity and both magnitudes are negative. 
 Note that we neglect the electron-like Fermi surface pocket near $\Gamma$ point originating from the $e_g$ orbital, which appears for  $\theta \geq 7^\circ$.  
 In the present study we concentrate the effect of the Fermi surface topology from the  $\gamma$ band. 
 
 The effective Hamiltonian is written as
\begin{eqnarray}
H&=&H_{\rm K}+H_{\rm int},\\
H_{\rm K}&=&-t\sum_{il\sigma} \left(c^{\dag}_{il\sigma}c_{i+1l\sigma} +c^{\dag}_{il\sigma}c_{il+1\sigma}+h.c.\right)\nonumber \\
&&-t'\sum_{il\sigma} \left(c^{\dag}_{il\sigma}c_{i+1l+1\sigma} +c^{\dag}_{il\sigma}c_{i-1l+1\sigma}+h.c.\right)\nonumber \\
&&-\mu\sum_{il\sigma}n_{il\sigma},\nonumber \\
H_{\rm int}&=&U_a\sum_{il\sigma\sigma'} \left(n_{il\sigma}n_{i+1l\sigma'} +n_{il\sigma}n_{il+1\sigma'}\right)
\nonumber \\
&+&V_a\sum_{il\sigma\sigma'} \left(n_{il\sigma}n_{i+1l+1\sigma'} +n_{il\sigma}n_{i-1l+1\sigma'}\right),
\end{eqnarray}
where $c_{il\sigma}$ ($c^{\dag}_{il\sigma}$) is the annihilation (creation) operator for $d_{xy}$ electrons on the site $i$, the leg $l$ with spin $\sigma$ (=$\uparrow$ or $\downarrow$). $n_{il\sigma}=c^{\dag}_{il\sigma}c_{il\sigma}$ represents the number operator. 

\begin{figure}[t]
\includegraphics[width=80mm]{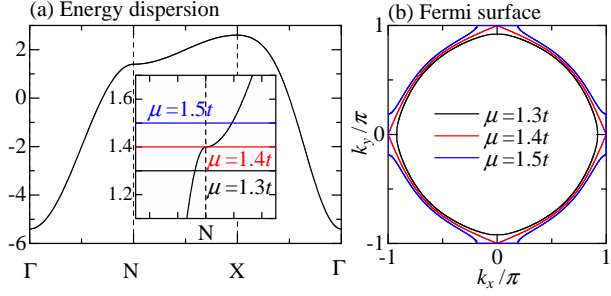}
\caption{(Color online) (a) Non-interacting energy dispersion in the bulk system obtained from the single band tight-binding model at $t=0.35t'$.  The inset shows the magnification near N point where the black, red and blue lines represent the Fermi level for the several choices of $\mu$. (b) Fermi surface for the several choices of $\mu$.}
\label{band_tba}
\end{figure}
The energy dispersion for the non-interacting ($U_a=V_a=0$) two-dimensional bulk system with the periodic boundary condition along the $x$ and $y$ directions is depicted in Fig. \ref{band_tba} (a), which roughly reproduces the $\gamma$ band in the DFT energy dispersion (Fig. \ref{band_LDA}). 
With increasing the chemical potential for the single band model, the Fermi surface topology changes at $\mu= 1.4t$ $(\equiv \mu_c)$ for $t'=0.35t$ shown in Fig. \ref{band_tba} (b), in which the Fermi surface yielded by the shift of the chemical potential qualitatively describes that topology in the system with the rotation of the RuO$_6$ octahedron. 
The topological properties do not depend on the details of the band structure, but only on the Fermi surface topology and the band structure is not sensitive to the presence of the RuO$_6$ rotation except $\Gamma$ and M 
points. 
In the present treatment instead of considering the RuO$_6$ rotation effect directly,  we incorporate the rotation angle through the increase of the chemical potential in our lattice model ignoring the doubling of the unit cell, for simplicity. 

The BCS-type mean-field approximation is applied to the attractive interaction term to generate the spin-triplet superconducting state, so that the gap functions are defined as
\begin{eqnarray}
\Delta^x_l&=&\frac{1}{2}\left( \langle c_{il\uparrow}c_{i+1l\downarrow}\rangle+\langle c_{il\downarrow}c_{i+1l\uparrow}\rangle\right)
\label{eqn:Delta_x},\\
\Delta^y_l&=&\frac{1}{2}\left( \langle c_{il\uparrow}c_{il+1\downarrow}\rangle+\langle c_{il\downarrow}c_{il+1\uparrow}\rangle\right),\\
\Delta^+_l&=&\frac{1}{2}\left( \langle c_{il\uparrow}c_{i+1l+1\downarrow}\rangle+\langle c_{il\downarrow}c_{i+1l+1\uparrow}\rangle\right),\\
\Delta^-_l&=&\frac{1}{2}\left( \langle c_{il\uparrow}c_{i-1l+1\downarrow}\rangle+\langle c_{il\downarrow}c_{i-1l+1\uparrow}\rangle\right),
\label{eqn:Delta_-}
\end{eqnarray}
which represents in-plane equal-spin pairings.  

While the gap functions are defined as eqs. (\ref{eqn:Delta_x})-(\ref{eqn:Delta_-}), $\Delta^\alpha_l$ ($\alpha=y,+,-$) is not symmetric with respect to the center of ribbon ($l=L/2$). Thus it is redefined as
\begin{eqnarray}
\Delta'^\alpha_l=\left\{
\begin{array}{cc}
\Delta^\alpha_l/2&(l=1)\\
\Delta^\alpha_{l-1}/2&(l=L)\\
(\Delta^\alpha_{l-1}+\Delta^\alpha_{l})/2&({\rm otherwise})\\
\end{array}
\right..
\end{eqnarray}

\subsection{Superconducting state}

The order parameters are determined self-consistently for $U_a = -2t$ and $t'=0.35t$ at absolute zero temperature. 
Since this amplitude gives rise to large amplitude of the gap function, the coherence length becomes short, i.e., only a few lattice
constants. Hence, the number of legs $L = 100$ is sufficient to ensure independent edge states at the two edges and the ribbon center displaying essentially bulk properties. 

\begin{figure}[tb]
\includegraphics[width=85mm]{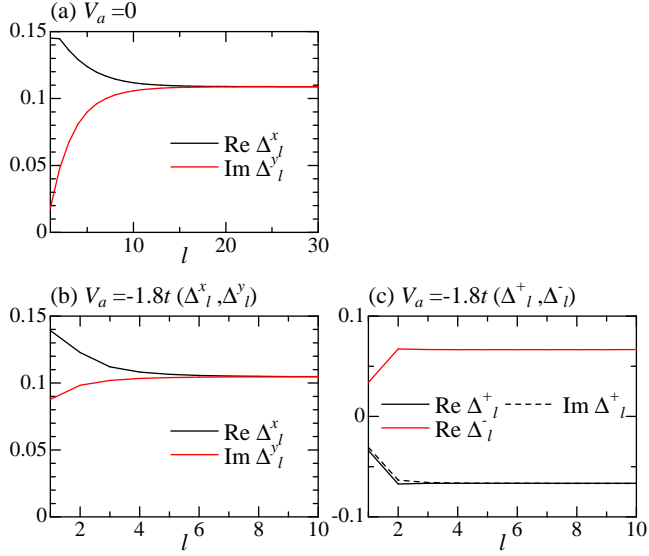}
\caption{(Color online) Gap functions as a function of leg $l$ for two choices of $V_a$ near an edge for $\mu=1.3t$.  }
\label{Delta_ribbon}
\end{figure}

First let us discuss the superconducting order parameters depicted in Fig. \ref{Delta_ribbon}. 
Apart from the vicinity of the edges, the real and imaginary parts of the gap functions have relative phase difference ($\pi/2$) and the same amplitudes, as follows
$\Delta^y=i\Delta^x$ and $\Delta^-=i\Delta^+$ with the angular momentum $L_z=+1$.
The latter is satisfied for the finite $V_a$, where the gap functions between nearest and next nearest neighbor sites have same chirality. 
Thus we find that the most stable pairing state has the chiral $p$-wave
form with $\bm d = \hat{z}(k_x + ik_y)$ for $V_a=0$, and $\bm d = \Delta \hat{z}(k_x + ik_y)+ \Delta' \hat{z}\{(k_x+k_y) + i(-k_x+k_y)\}$ for $V_a=-1.8t$ avoiding nodes in the excitation gap in a wide region of $\mu$ with $ \arg{(\Delta'/\Delta)} = \pi/4 $ in the bulk.
Note that although the pairing state with the angular momentum $L_z=-1$ is degenerate energetically and corresponds to the time-reversed state $\bm d = \Delta \hat{z}(k_x - ik_y)+ {\Delta'}^*  \hat{z}\{(k_x+k_y) - i(-k_x+k_y)\}$, we discuss mainly the  $L_z=+1$ state.

On the other hand, the $y$ ($x$) component of the gap function is suppressed (enhanced slightly) at the edge, which is attributed to the reflection of the Cooper pairs at the edges because the $y$ ($x$) component of the order parameter is odd under reflection at the edges.

\begin{figure}[tb]
\includegraphics[width=85mm]{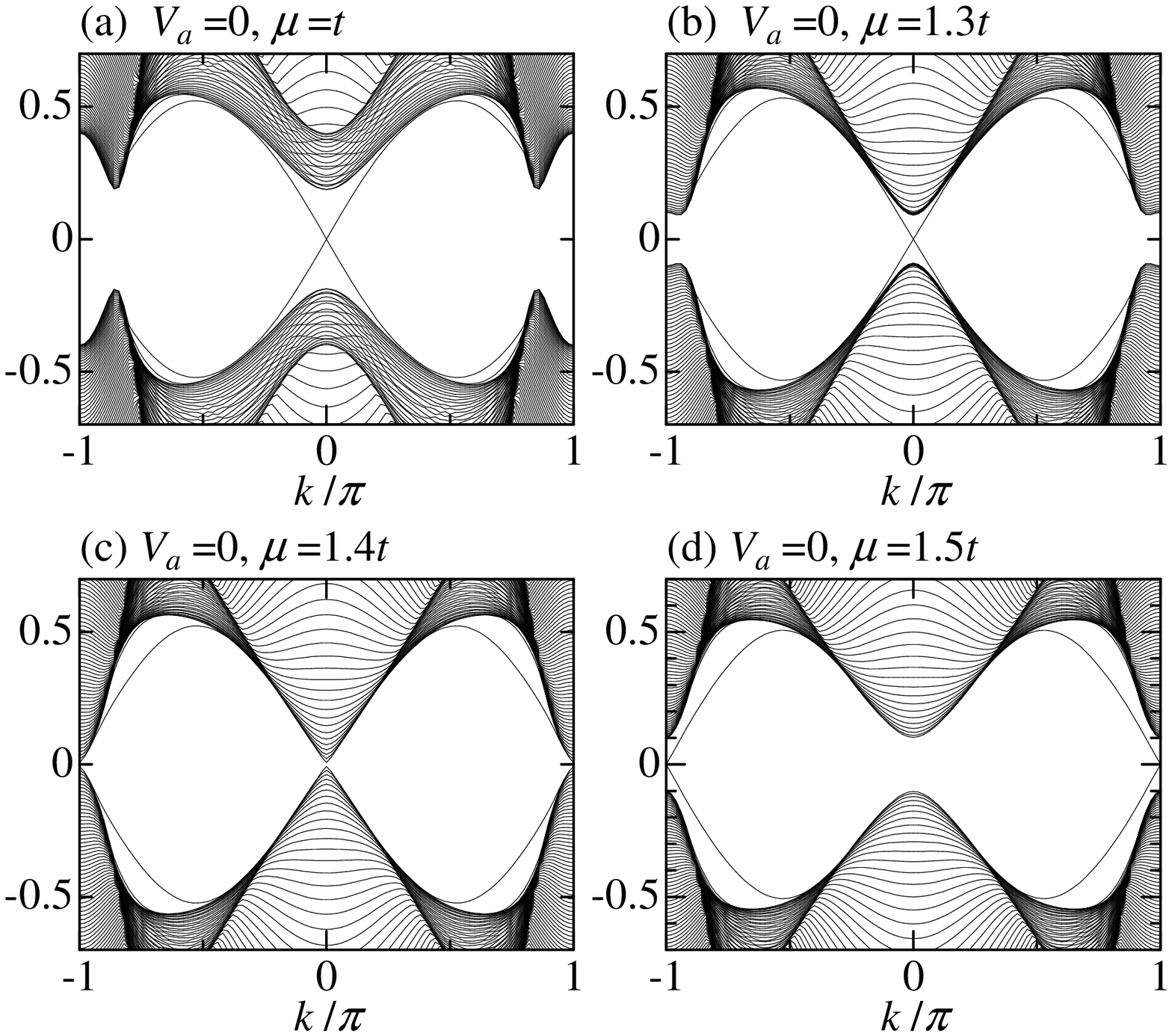}
\includegraphics[width=85mm]{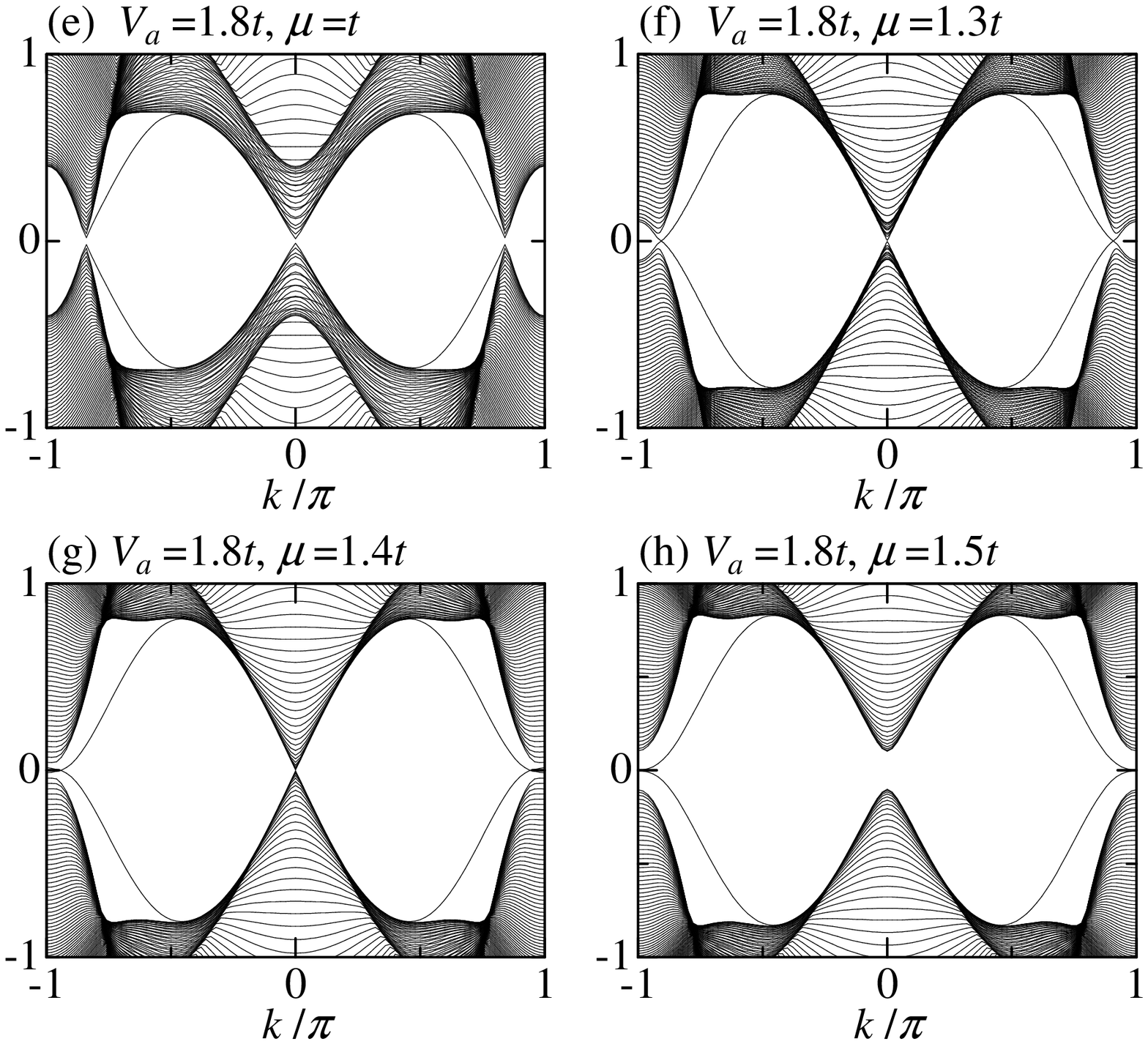}
\caption{Energy dispersions with open boundary condition along the $y$ direction at low energy sector; (a)-(d) $V_a=0$ and (e)-(h) $V_a=-1.8t$ for several choices of $\mu$.   }
\label{band_ribbon}
\end{figure}

In order to study the low-energy edge states, we show the energy dispersions for the ribbon model in Fig. \ref{band_ribbon}. 
Note here that we always assume translational invariance along the $x$ direction where the Fourier transition only in the $x$ direction is defined as
\begin{eqnarray}
c_{il\sigma}=\frac{1}{\sqrt{L_x}}\sum_{k}c_{kl\sigma} e^{-ik x_i},
\end{eqnarray}
where $k$ stands for the momentum along the $x$ direction. $x_i$ is the $x$ coordinate of site $(i,l)$ and $L_x$ is the number of the sites in the $x$ direction. 

The subgap states appear in the quasiparticle excited gap.  
Their spectrum shows a strong linear dispersions at $k \sim 0$ and/or $\sim \pm\pi$, which are the chiral edge states resulting from the $\gamma$ band whereby each edge contributes one of the two chiral branches which is topologically protected.

The chiral edge state crossing zero-energy shifts from $k=0$ to $k=\pm \pi$ around $\mu=\mu_c$, 
which indicates the Lifshitz transition of the $\gamma$ band from electron-like ($\mu<\mu_c$) to hole-like ($\mu>\mu_c$).  
For $V_a=-1.8t$, there are three crossing points at the low energy region slightly below $\mu_c$ (shown in Fig. \ref{band_ribbon} (f)). 
Compared with the $V_a=0$ case, since the superconducting gap around $k=0$ and $k=\pm \pi$ is reduced in the wide range of $\mu$, 
the edge states may become rather fragile for larger $V_a$ region.

\subsection{Currents near the edges}

Here, let us discuss the transport property near the edges. 
The broken time-reversal symmetry realized in the chiral $p$-wave superconducting phase generates the circular current along the edges even without external field. 
The expectation value of the current along the $x$ direction is defined as
\begin{eqnarray}
\langle \hat{j} \rangle &=&\left\langle\frac{\partial H_{\rm K}}{\partial k}\right\rangle
=\sum_{\sigma}\left\{\sum_{l=1}^{L}\langle j^{(1)}_{l\sigma}\rangle+\sum_{l=1}^{L-1}\langle j^{(2)}_{l\sigma}\rangle\right\}, \\
j^{(1)}_{l\sigma}&=&\frac{1}{L_x}\sum_{k}(2t\sin k)  c^{\dag}_{kl\sigma}c_{kl\sigma}, \\
j^{(2)}_{l\sigma}&=&\frac{1}{L_x}\sum_{k}(2it'\sin k) \left( c^{\dag}_{kl\sigma}c_{kl+1\sigma}+c^{\dag}_{kl+1\sigma}c_{kl\sigma}\right),  
\end{eqnarray}
where $L_x$ is the number of sites in the $x$ direction.
For the present single band model, the spin index can be omitted in the current. 
Since $j^{(2)}_{l\sigma}$ is not symmetric with respect to the center of ribbon ($l=L/2$), we redefine it as
\begin{eqnarray}
j^{(2)'}_{l\sigma} =\left\{
\begin{array}{cc}
j^{(2)}_{l\sigma} /2&(l=1)\\
j^{(2)}_{l-1\sigma} /2&(l=L)\\
\left( j^{(2)}_{l-1\sigma} + j^{(2)}_{l\sigma}\right)/2&({\rm otherwise})\\
\end{array}
\right..
\end{eqnarray}
The observable current is given as the summation of the current flowing at each leg. 
The net current and the $l$-dependent current along an edge are given as
\begin{eqnarray}
\langle \hat{j} \rangle_{\rm net}&=&\sum_{l=1}^{L/2}\langle \tilde{j}_l \rangle, \label{jnet} \\
\langle \tilde{j}_l \rangle&=&\langle j^{(1)}_{l\sigma}\rangle+\langle j^{(2)'}_{l\sigma}\rangle.
\end{eqnarray}

\begin{figure}[tb]
\includegraphics[width=85mm]{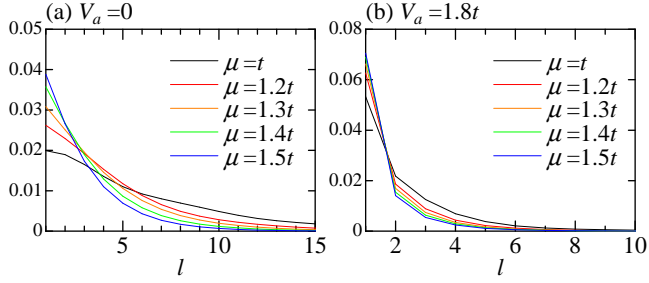}
\caption{(Color online) $l$-dependent currents for the various choices of $\mu$ near an edge; (a) $V_a=0$ and (b) $V_a=-1.8t$.  }
\label{current}
\end{figure}

Figure \ref{current} shows the $l$-dependent currents. 
With decreasing the distance from the edge, the $l$-dependent current amplitude decreases monotonically. 
We stress that the currents have monotonous change around $\mu=\mu_c$ in both $V_a=0$ and $V_a=-1.8t$ cases.  
The chirality switches at the Lifshitz transition and the carriers run in the opposite direction in the edge state. However, at the same time the charge changes sign as well so that the charge current is always in the same direction. 

The next nearest neighbor interaction enhances (suppresses) the $l$-dependent current at $l=1$ (otherwise) because the interaction enlarges the superconducting gap except $k\sim 0$ and $k\sim \pm \pi$, which leads to the appearance of larger dispersive subgap bands.  
However, the spatial dependence of the current  varies only weakly as a function of $\mu$, because in the larger $V_a$ region, the superconducting gap around $k=0$ and $k\sim \pm \pi$ is suppressed for a wide range of $\mu$ and the subgap dispersion is insensitive to the change of $\mu$ around $\mu=\mu_c$, shown in Fig.\ref{band_ribbon}. 

\begin{figure}[tb]
\includegraphics[width=60mm]{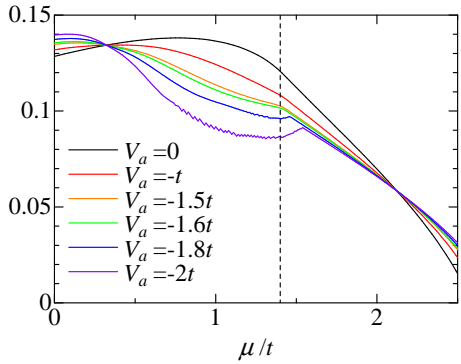}
\caption{(Color online) Net current as a function of $\mu$ for several choices of $V_a$.  The dashed line stands for $\mu=\mu_c$.  }
\label{jsum}
\end{figure}
The net edge current $\langle \hat{j} \rangle_{\rm net}$ as a function of $\mu$ is displayed in Fig.\ref{jsum}, defined in Eq.(\ref{jnet}).
The variation of  $\langle \hat{j} \rangle_{\rm net}$ as a function of $V_a$ is large for $ \mu < \mu_c $, while in the range $ \mu > \mu_c $ we find only moderate dependence on $V_a$. It is also obvious that there is not dramatic change of the net current when passing through the Lifshitz transition.  

\subsection{Topological Property}

The edge states of the materials with fully opened insulating or superconducting gap are strongly related to the topological property of the bulk system. 
By means of the two-dimensional bulk model, we investigate the topological property, where the mean-field Hamiltonian reads
\begin{eqnarray}
H^{\rm MF}=\sum_{\bm k}(c^{\dag}_{\bm k\uparrow},c_{-\bm k\downarrow})
\left(
\begin{array}{cc}
\varepsilon_{\bm k}  & \Delta_{\bm k}\\
\Delta^*_{\bm k} & -\varepsilon_{\bm k}
\end{array}
\right)
\left(
\begin{array}{c}
c_{\bm k\uparrow}\\
c^{\dag}_{-\bm k\downarrow}
\end{array}
\right), 
\end{eqnarray}
where $c_{\bm k\sigma}$ ($c^{\dag}_{\bm k\sigma}$) is the annihilation (creation) operator for electrons on the momentum $\bm k$ with spin $\sigma$ (=$\uparrow$ or $\downarrow$).
$\varepsilon_{\bm k}$ is the one-particle energy dispersion written as $\varepsilon_{\bm k}=-2t(\cos k_x+\cos k_y)-4t'\cos k_x \cos k_y-\mu$, and  
$\Delta_{\bm k}$ represents the gap function for the spin-triplet sector defined as
\begin{eqnarray}
\Delta_{\bm k}&=& 2iU_a\left(\Delta^x\sin k_x+\Delta^y\sin k_y\right)\nonumber \\
&+&2iV_a\left\{\Delta^+\sin (k_x+k_y)+\Delta^-\sin (-k_x+k_y)\right\}. 
\end{eqnarray}
While the gap functions are defined in eqs. (\ref{eqn:Delta_x})-(\ref{eqn:Delta_-}), those in the two-dimensional bulk system are independent of the leg index $l$ because of the presence of the translational invariance along the $x$ and $y$ directions. 
Note that the obtained gap functions correspond to those at the center of the ribbon model . 

The topological property with the fully gaped system is characterized the so-called Chern number~\cite{volo97,furu01}, which is defined as 
\begin{eqnarray}
N_{\rm c}=\frac{1}{4\pi}\int \,dk_x dk_y\>\hat{\bm d}_{\bm k}\cdot \left( \frac{\partial \hat{\bm d}_{\bm k}}{\partial k_x}\times \frac{\partial \hat{\bm d}_{\bm k}}{\partial k_y} \right),
\end{eqnarray}
where 
\begin{eqnarray}
&&{\bm d}_{\bm k}=({\rm Re}\Delta_{\bm k},{\rm Im}\Delta_{\bm k},\varepsilon_{\bm k}),\\
&&\hat{\bm d}_{\bm k}={\bm d}_{\bm k}/|{\bm d}_{\bm k}|,\>\>|{\bm d}_{\bm k}|=\sqrt{|\Delta_{\bm k}|^2+\varepsilon_{\bm k}^2}.
\end{eqnarray}
It is easily derived from the linear response theory for the single band model~\cite{thou82,kohm85}. 

\begin{figure}[tb]
\includegraphics[width=85mm]{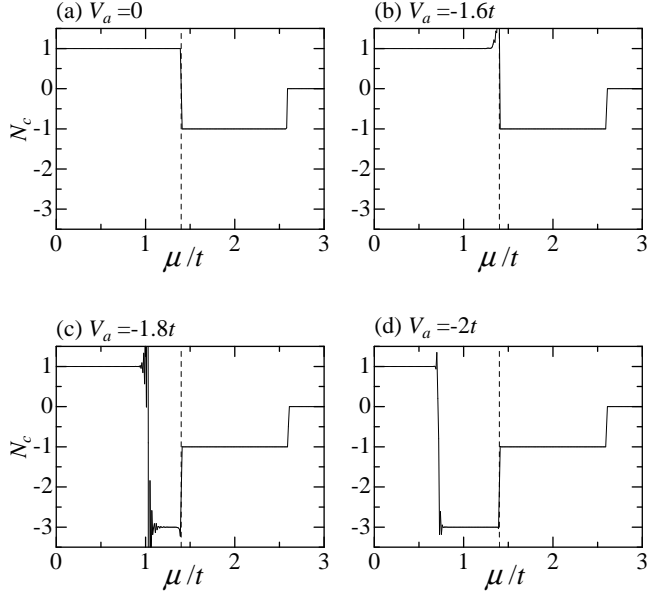}
\caption{The Chern number as a function of $\mu$ for the several choices of $V_a$.  The dashed line stands for $\mu=\mu_c(=1.4t)$. }
\label{nc}
\end{figure}

The resulting Chern number is shown as a function of $\mu $ in Fig. \ref{nc}. 
Note that $N_{\rm c}$ deviates from integer at some values of $\mu$ where the Chern number ''jumps'' between different values. These fluctuations originate from the closing of the gap at these points, making the Chern number ill-defined. 

$N_{\rm c}$ for $\mu<\mu_c$ depends on the amplitude of $V_a$ where $N_{\rm c}=1(-3)$ for smaller (larger) $V_a$. 
With increasing $V_a$, the $N_{\rm c}=-3$ region concerning $\mu$ is enlarged. 
Since the chemical potential of tetragonal Sr$_2$RuO$_4$ is slightly lower than $\mu=\mu_c$, the topological property of bulk Sr$_2$RuO$_4$ depends strongly on $V_a$. 
On the other hand, we find $N_{\rm c}=-1$ for $\mu>\mu_c$, which shows that the critical value $\mu_c$ at the Lifshitz transition is independent of the amplitude of $V_a$. 
Note that in the larger-$V_a$ region, the topological property near the Lifshitz transition depends not on $U_a$ but on $V_a$, which leads to the topological number $N_{\rm c}=-3$ $(-1)$ for $\mu<\mu_c$ $(\mu>\mu_c)$ in the case of $U_a\rightarrow 0$ and $V_a \ne 0$. 

\begin{figure}[tb]
\includegraphics[width=60mm]{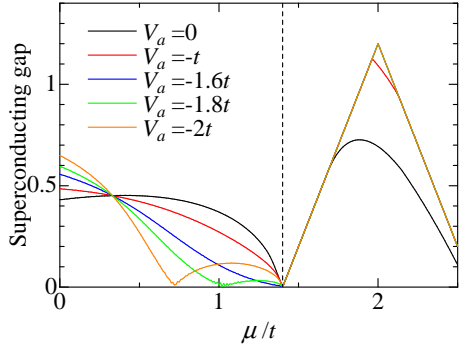}
\caption{(Color online) The minimum of the superconducting gap as a function of $\mu$ for the several choices of $V_a$. The dashed line represents the $\mu=\mu_c$.  }
\label{supergap}
\end{figure}

The results concerning Chern number show that the topological properties change at the Lifshitz transition, where also the superconducting gap vanishes at the van Hove points. This feature is independent of the pairing interaction. 
Since the Fermi level of tetragonal Sr$_2$RuO$_4$ is slightly below $\mu_c$, the minimum of the superconducting gap of Sr$_2$RuO$_4$ in the whole Brillouin Zone is very small, as depicted in Fig. \ref{supergap}.
In particular, the superconducting gap is rather small over a wide region around $\mu=\mu_c$ for larger $V_a$. 
This result indicates that for larger $V_a$, the gap is very small not only near the surface but also in the depth of the material. 
At $\mu>\mu_c$, the next nearest neighbor interaction does not affect the amplitude of the superconducting gap. 
However, since even near the surface the chemical potential $\mu$ is almost equal to or slightly larger than $\mu_c$, the gap has also very small amplitude. 
The topological superconducting states are realized at each layer in both surface and bulk regions, and the topological number near the surface is different from that of the bulk system. However, the topological protection is not robust and the presence of  disorder such as impurity and defect may affect the topological property (see also Ref.\ref{wang13}).

\section{Summary and discussion}
We have investigated the topological and transport properties of the spin-triplet chiral $p$-wave superconductor Sr$_2$RuO$_4$ with electronic reconstruction by doubling of the unit cell near the surface. 

The detailed energy band structures have been obtained using the first principles calculation. 
With increasing the rotation angle which means to access the surface, the Fermi surface originating from the $\gamma$ band clearly shows the Lifshitz transition. 

We also discuss the topological and transport properties with the lattice fermion model extracting the $\gamma$ band based on the DFT calculation.  
While the topological property strongly depends on the rotation angle and the amplitude of the nearest neighbor interaction in the bulk Sr$_2$RuO$_4$, 
the critical amplitude of the chemical potential at the Lifshitz transition and the topological property near the surface are almost independent of the presence of the nearest neighbor interaction
where the edge current changes monotonously around the Lifshitz transition. 
However, in the larger nearest neighbor interaction region, the superconducting gap is strongly reduced not only near the surface but also bulk system.  
The topological protection to the edge state is not robust and disorder may affect seriously in the transport property. 
This suppresses the edge current, which may be the one of the reason concerning the discrepancy between theoretical and experimental studies concerning the detect of the edge current.

\section*{Acknowledgment}
We are grateful to A. Bouhon, T.M. Rice, J. Goryo and T. Saso for helpful discussions. 
This work was financially supported by the Ministry of Education, Culture, Sports, Science and Technology, Japan.
YI is grateful for hospitality during his visit to the Pauli Center for theoretical studies of ETH Zurich.



\end{document}